\documentstyle[multicol,psfig,aps,prd]{revtex}
\begin{document}
\newcommand{\gtsima}{$\; \buildrel > \over \sim \;$}
\newcommand{\ltsima}{$\; \buildrel < \over \sim \;$}
\newcommand{\simgt}{\lower.5ex\hbox{\gtsima}}
\newcommand{\simlt}{\lower.5ex\hbox{\ltsima}}
\newcommand{\himpc}{{\hbox {$h^{-1}$}{\rm Mpc}} }
\newcommand{\bfq}{{\mbox{\boldmath $q$}}}
\newcommand{\bfx}{{\mbox{\boldmath $x$}}}
\newcommand{\bfv}{{\mbox{\boldmath $v$}}}
\newcommand{\bfpsi}{{\mbox{\boldmath $\psi$}}}
\newcommand{\bfPsi}{{\mbox{\boldmath $\Psi$}}}
\newcommand{\mPsi}{{\mit\Psi}}
%
\title{\large\bf  BEYOND ZEL'DOVICH-TYPE APPROXIMATIONS IN GRAVITATIONAL
INSTABILITY THEORY\\
--- Pad\'e Prescription in Spheroidal Collapse --- }
\author{Takahiko Matsubara\thanks{Electronic address:
matsu@phys.s.u-tokyo.ac.jp}}
\address{Department of Physics, The University of Tokyo, Hongo,
Bunkyo-ku, Tokyo 113, Japan.;\\ Research Center for the Early Universe,
Faculty of Science, The University of Tokyo, Tokyo 113, Japan.} 
\author{Ayako Yoshisato\thanks{Electronic address:
ayako@cosmos.phys.ocha.ac.jp} \and Masahiro Morikawa\thanks{Electronic 
address: hiro@phys.ocha.ac.jp}} 
\address{Department of Physics, Ochanomizu University, 2-1, Otsuka 2,
Bunkyo-ku 112, Japan.} 

\maketitle
\begin{abstract}
Among several analytic approximations for the growth of density
fluctuations in the expanding Universe, Zel'dovich approximation in
Lagrangian coordinate scheme is known to be unusually accurate even in
mildly non-linear regime.  This approximation is very similar to the
Pad\'e approximation in appearance.  We first establish, however, that
these two are actually different and independent approximations with
each other by using a model of spheroidal mass collapse.  Then we
propose Pad\'e-prescribed Zel'dovich-type approximations and
demonstrate, within this model, that they are much accurate than any
other known nonlinear approximations.
\end{abstract}

\begin{multicols}{2}
  When we analyze the growth of density fluctuations in the expanding
  Universe by analytical methods, the Zel'dovich-type approximations
  (ZTA hereafter) are known to be unusually accurate even in mildly
  non-linear regime for unknown reason%
~\cite{Zel'dovich1,Zel'dovich2,Munshi,Sahni1,Sahni2,Buchert1,Melott,Bouchet}.
  These Zel'dovich-type approximations are grounded on the Lagrangian
  coordinate scheme and are one-dimensional-exact; they become exact
  in the plain parallel mass distributions.  The validity of ZTA has
  been argued recently based on these physical
  properties~\cite{Yoshisato}.  On the other hand, the appearance of
  these ZTA are very similar to the rational expansion method called
  Pad\'e approximations.  Though they have been widely used in the
  literature, the validity of these approximations has not yet been
  established as well.

  We would like first to compare these Zel'dovich-type and Pad\'e
  approximation methods. It is almost impossible and is not pragmatic
  to argue in general analytic form.  Therefore in this letter, we
  restrict our consideration to a model of spheroidal collapse which
  we can solve semi-analytically.  If the above two types of
  approximations are actually the same, then they would give the same
  result for this restricted model.  We demonstrate, in the first part
  of this letter, that this is not the case and conclude that they are
  independent approximation schemes.  Then this fact suggests a
  possibility to go beyond ZTA by Pad\'e prescription on ZTA.  We
  demonstrate, in the second part of this letter, that this Pad\'e
  prescription dramatically improves ZTA.

  In the gravitational instability theory, the non-relativistic matter
  with zero pressure in an Einstein-de Sitter (EdS) universe is
  described by the following set of equations (see
  Ref.~\cite{Peebles}),
\begin{eqnarray}
   &&
   \dot{\delta} + \nabla\cdot[(1 + \delta)\bfv] = 0,
   \label{eq1}
   \\
   &&
   \dot{\bfv} + 2 H \bfv + (\bfv\cdot\nabla)\bfv +
   \nabla\Phi = \bf0,
   \label{eq2}
   \\
   &&
   \nabla^2 \Phi = \frac32 H^2 \delta,
   \label{eq3}
\end{eqnarray}
where $\bfx$, $\bfv(\bfx,t)$, $\Phi(\bfx,t)$ are respectively
position, peculiar velocity, peculiar potential in comoving
coordinate.
The scale factor $a$ varies as $a \propto t^{2/3}$ and the Hubble
parameter is $H = \dot{a}/a = 2/(3t)$.  Although we consider only EdS
universe for simplicity in this letter, it is straightforward to
generalize the analyses in general Friedman-Lem\^{\i}tre universes.

In the Eulerian coordinate scheme, the linear solution of them has a
simple form $\delta_{\rm L} (\bfx, t) =\delta_{\rm in} (\bfx) a
(t)/a_{\rm in}$, neglecting the decaying mode.  Considering this to be
a small parameter, we can naturally expand the full solution in powers
of $\delta_{\rm L}$: $\delta = \delta_{\rm L} + \delta^{(2)} +
\delta^{(3)} + \cdots$, $\Phi = \Phi_{\rm L} + \Phi^{(2)} + \Phi^{(3)}
+ \cdots$, $\bfv = \bfv_{\rm L} + \bfv^{(2)} + \bfv^{(3)} + \cdots$,
where $\delta^{(n)}$, $\Phi^{(n)}$, $\bfv^{(n)}$ are assumed to be of
order $(\delta_{\rm L})^n$. (e.g., see~\cite{Peebles,Fry,Goroff}).

In the Zel'dovich-type
approximations~\cite{Zel'dovich1,Munshi,Bouchet,Buchert2,Buchert3,Catelan},
we work in the Lagrangian coordinate scheme in which the location of a
mass element $\bfx$ of the fluid is expressed by the initial location
$\bfq$ and the time dependent displacement vector $\bfPsi$ as
$\bfx(\bfq,t) = \bfq + \bfPsi(\bfq,t)$.  Then the density contrast
$\delta[\bfx(\bfq,t),t] = \det[\partial x_i/\partial q_j]^{-1} - 1$ is
determined by solving the equation of motion
\begin{eqnarray}
&&
   \left[\frac{d^2\mPsi_{i,j}}{dt^2} +
     2 H \frac{d\mPsi_{i,j}}{dt} \right]\left(J^{-1}\right)_{ji} +
   \frac32 H^2 \left(J^{-1} - 1\right) = 0,
   \label{eq2-la4} 
\\
&&
   \epsilon_{ijk} \frac{d\mPsi_{j,l}}{dt}
   \left(J^{-1}\right)_{lk} = 0,
   \label{eq2-la6}
\end{eqnarray}
where $d/dt$ is the Lagrangian time derivative, $J_{ij} = \partial
x_i/\partial q_j = \delta_{ij} + \mPsi_{i,j}$, $J = \det J_{ij}$.
This Eq.~(\ref{eq2-la4}) is obtained from Eqs.~(\ref{eq2}) and
(\ref{eq3}), and Eq.~(\ref{eq2-la6}) corresponds to the usual Eulerian
vorticity-free condition. These nonlinear equations for $\bfPsi$ can
be solved by the method of iteration considering
$\partial\mPsi_i/\partial q_j$ as small expansion parameter:
$\mPsi_{i,j} = \mPsi_{i,j}^{(1)} + \mPsi_{i,j}^{(2)} +
\mPsi_{i,j}^{(3)} + \cdots$.  In EdS universe, the time dependence of
each terms is separated from its spatial dependence: $\bfPsi^{(n)} =
(2/(3 a^2 H^2))^n \bfpsi^{(n)}(\bfq)$. The first-order solution
$\psi^{(1)}_i = - \partial_i \phi_{\rm L} (\bfq)$ is the original
Zel'dovich approximation (ZA).

Now we introduce a model of collapsing homogeneous ellipsoid. We
parameterize the density perturbation $\delta(\bfx,t)$ as
\begin{eqnarray}
  \delta(\bfx, t) =
  \delta_{\rm e}(t)\,\,\,
  \Theta\left(
    1 - 
    \frac{x_1^{\,2}}{\alpha_1^{\,2}(t)} -
    \frac{x_2^{\,2}}{\alpha_2^{\,2}(t)} -
    \frac{x_3^{\,2}}{\alpha_3^{\,2}(t)}
  \right),
  \label{eq:m3}
\end{eqnarray}
where $\alpha_i$ are the half-length of the principal axes of the
ellipsoid and $\Theta$ is the step function.  The solution of the
Poisson Eq.~(\ref{eq3}) inside this homogeneous ellipsoid is known
(see, Ref.~\cite{Kellogg})
and it becomes $\Phi = \frac{3}{8} H^2 \delta_{\rm e}
  \sum_{i=1}^3 A_i(t) x_i^{\,2}$,
and the equations of motion for three
$\alpha_i$ are given by~\cite{Yoshisato,WS}
\begin{eqnarray}
  \ddot{\alpha}_i + 2H \dot{\alpha}_i =
  - \frac{3}{4} H^2 \delta_{\rm e} A_i \alpha_i,
  \label{eq:m9}
\end{eqnarray}
where
\begin{eqnarray}
  A_i = \alpha_1\alpha_2\alpha_3
  \int_0^\infty (\alpha_i^{\,2} + \lambda)^{-1}
  \prod_{j=1}^3
  (\alpha_j^{\,2} + \lambda)^{-1/2} d\lambda.
  \label{eq:m5}
\end{eqnarray}
The density contrast of the ellipsoid is obtained by observing that $1
+ \delta$ is inversely proportional to $\alpha_1 \alpha_2 \alpha_3$.
These equations are solved by numerical integration. The solutions are
'semi-analytic' in this sense. In the following, we only consider
spheroidal case, $\alpha_1 = \alpha_2$ for simplicity.

For the system of spheroidal perturbations, we apply ZTA, resulting
in~\cite{Yoshisato}
\begin{eqnarray}
& &_{} 
\psi^{(1)}_{(1,2)} = \mp \displaystyle\frac{a^3 H^2 q_{(1,2)}}{2}
  (1 + h_{\rm in}) , 
\label{1sted12}\\
& &_{} 
\psi^{(1)}_3 = \mp \displaystyle\frac{a^3 H^2 q_3}{2} (1 - 2 h_{\rm in}) ,
\label{1stelidis} \\
& &_{} 
\psi^{(2)}_{(1,2)}
= - \displaystyle\frac{3 a^6 H^4 q_{(1,2)}}{28} (1 + h_{\rm in}
- h_{\rm in}^2 - h_{\rm in}^3), \\
& &_{} 
\psi^{(2)}_3
= - \displaystyle\frac{3 a^6 H^4 q_3}{28} (1 - 2 h_{\rm in}
- h_{\rm in}^2 + 2 h_{\rm in}^3), 
\label{2ndelidis} \\
& &_{} 
\psi^{(3)}_{(1,2)} =
\mp \displaystyle\frac{a^9 H^6 q_{(1,2)}}{504} ( 23 + 23  h_{\rm in}
- 39  h_{\rm in}^2 \nonumber\\
& &_{} 
\qquad\qquad\qquad\qquad\qquad
 - 25  h_{\rm in}^3 + 44 h_{\rm in}^4 + 30  h_{\rm in}^5 ), \\
& &_{} 
\psi^{(3)}_3 =
\mp \displaystyle\frac{a^9 H^6 q_3}{504} ( 23 - 46  h_{\rm in}
- 39  h_{\rm in}^2  \nonumber\\
& &_{} 
\qquad\qquad\qquad\qquad
 + 92  h_{\rm in}^3 + 2 h_{\rm in}^4 - 60  h_{\rm in}^5 ),
\label{3rdelidis}
\end{eqnarray}
where we have changed the parametrization, $A_1 = A_2 = \frac{2}{3} (1
+ h)$, $A_3 = \frac{2}{3} (1 - 2 h)$~\cite{foot1}, and $h_{\rm in} = h
(t_{\rm in})$, with $t_{\rm in}$ being the initial time for numerical
integration.  The density contrast in these ZTA is given by
\begin{eqnarray}
\delta = \frac{q_1 q_2 q_3}{(q_1 + \mPsi_1)(q_2 + \mPsi_2)(q_3 + \mPsi_3)}
 - 1, \label{yuragi}
\end{eqnarray}
where $\mPsi_{i} = \mPsi^{(1)}_{i}$ corresponds to the original ZA,
$\mPsi_{i} = \mPsi^{(1)}_{i} + \mPsi^{(2)}_{i}$ corresponds to
post-Zel'dovich approximation (PZA), $\mPsi_{i} = \mPsi^{(1)}_{i} +
\mPsi^{(2)}_{i} + \mPsi^{(3)}_{i}$ corresponds to post-post-Zel'dovich
approximation (PPZA).

In contrast to the above Lagrangian perturbation methods, the surface
of the spheroid cannot be explicitly expressed in Eulerian
perturbation methods. We simply transform the expression already
obtained in Lagrangian perturbation scheme to that in Eulerian
perturbation scheme.  This is based on the fact that the small
expansion parameters $\mPsi_{i,j}$ in Lagrangian scheme and $\delta$ in
Eulerian scheme are the same order and thus $\mPsi^{(n)}_{i,j} \ \sim
\delta^{(n)}$.  In our case, $\mPsi^{(n)} \propto a^n$, so Eulerian
perturbative series can be simply obtained by expanding equation
(\ref{yuragi}) in terms of expansion factor $a$:
\begin{eqnarray}
&&
\delta = \pm a + \left( \frac{17}{21} +
 \frac{4}{21} h_{\rm in}^2\right) a^2
\nonumber\\
&&
\qquad
\pm \left(\frac{341}{567} + \frac{74}{189} h_{\rm in}^2 -
  \frac{4}{81} h_{\rm in}^3 - \frac{8}{189} h_{\rm in}^4\right) a^3.
\label{eulerexp}
\end{eqnarray}
\par
Now we introduce Pad\'{e} approximation associated with the
Perturbative expansions in Eulerian coordinate scheme. Pad\'{e}
approximation of type-$(M,N)$ for a given unknown function $f(x)$ is
expressed as the ratio of two polynomials.
\begin{eqnarray}
  f_{\mbox{\scriptsize Pad\'e}(M,N)} \equiv \left(\displaystyle
    \sum_{k = 0}^{M} a_k x^k\right) \left(1 + \displaystyle \sum_{k =
    1}^{N} b_k x^k \right)^{-1}.
\end{eqnarray}
The constant parameters $a_k$ and $b_k$ are determined so that they
maximally yield the known Taylor expansion of the function
$f(x)$~\cite{Bender} up to $(M + N)$-th order.  The density contrast
in the spheroidal model up to the third-order perturbation is given by
Eq.~(\ref{eulerexp}). The corresponding Pad\'{e} approximation of
type-(1,2) is given by
\begin{eqnarray}
&&
\delta = \pm a 
\left[ 1 \mp \displaystyle\frac{17 + 4 h_{\rm in}^2}{21} a \right.
\nonumber\\
&&
\quad
\left.
+ \left(\displaystyle\frac{214}{3969} - \frac{110}{1323} h_{\rm in}^2
+ \frac{4}{81} h_{\rm in}^3 + \frac{104}{1323} h_{\rm in}^4 \right)
a^2 \right]^{-1} .
\label{eq:mm2}
\end{eqnarray}
We observe that the ZTA density contrast of Eq.~(\ref{yuragi}) and
Pad\'e density contrast of Eq.~(\ref{eq:mm2}) are definitely different
with each other.  Actually the plots of these approximations in Fig.~1
for spherical perturbations clearly demonstrate the difference.  Some
of lines in Fig.~1 have been previously appeared~\cite{Munshi,Sahni2}.

\end{multicols}
\begin{figure}[t]
\begin{center}
   \leavevmode\psfig{figure=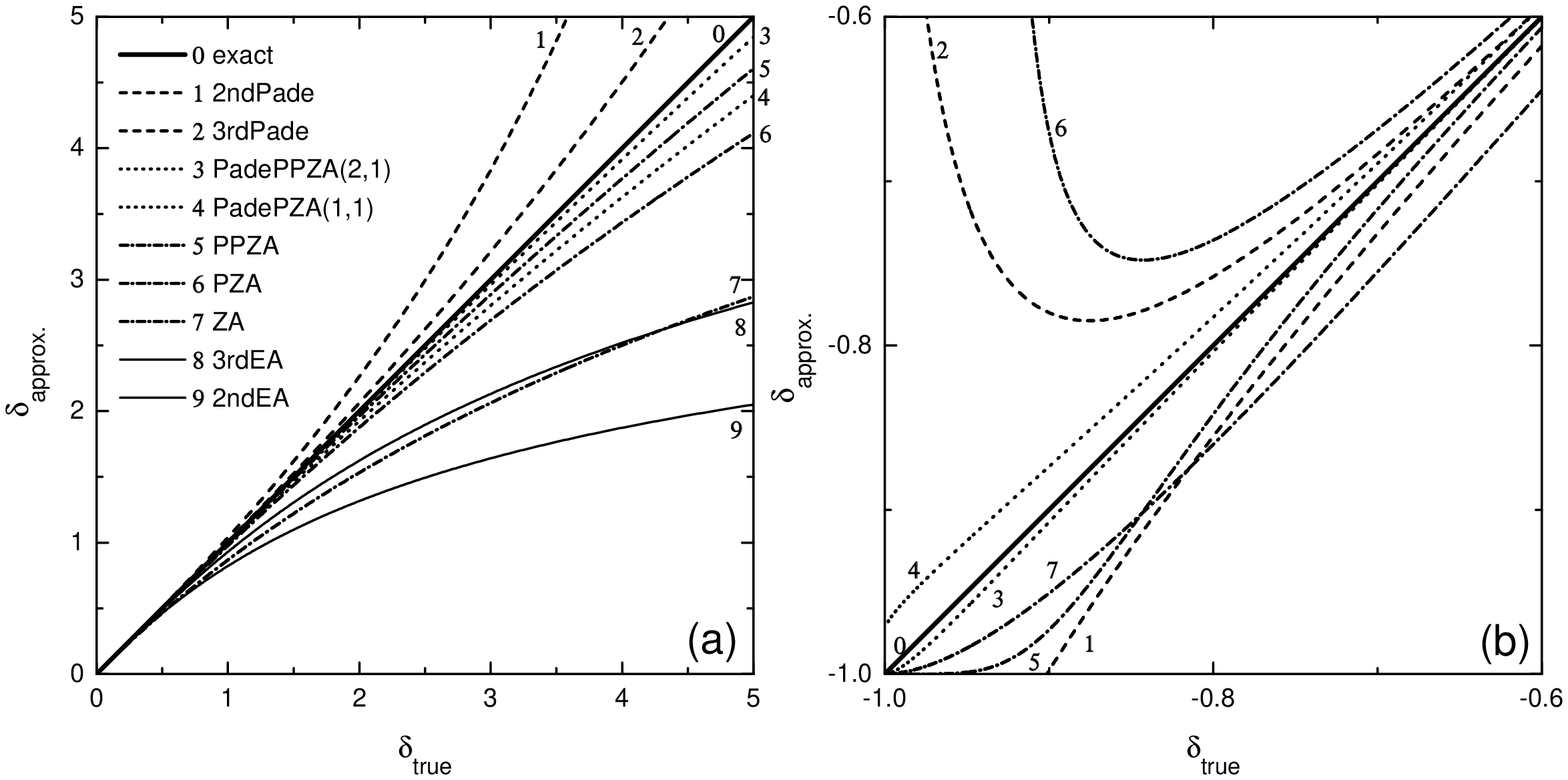,width=13cm,angle=270}
\end{center}
\caption{ Density contrasts in Eulerian approximations and
ZTA and their corresponding Pad\'e approximations in the model of
spherical density perturbations.  Vertical axis is the density
contrasts $\delta_{\rm approx}$ of each approximations and the
horizontal axis is the true solution $\delta_{\rm true}$.  Fig. 1(a)
is for a case of positive perturbation and Fig. 1(b) is for a case of
negative perturbation.}
\end{figure}
\begin{multicols}{2}

  Since the Zel'dovich-type approximations and Pad\'e approximations
  are different scheme with each other, we may be able to invent
  better approximations by combining these two scheme.  In doing so,
  we notice that the each factor $\mPsi_i$ in
  Eqs.~(\ref{1sted12})-(\ref{3rdelidis}) has an appropriate series
  expansion form.  Therefore we apply Pad\'e approximation for these
  factors $\mPsi_i$ separately.  For example, Pad\'e approximation for
  the factor $\mPsi_1$ becomes
\begin{eqnarray}
& &_{} 
\mPsi_{1 \, \mbox{\scriptsize Pad\'e}(1,2)} = 
\mp \frac{a (1 + h_{\rm in}) q_1}{3}
\left[1 \mp \displaystyle\frac{a (1 - h_{\rm in}^2)}{7}\right.
\nonumber\\
&& \qquad\qquad \left.
 - \displaystyle\frac{a^2(80 - 111 h_{\rm in}^2 + 98 h_{\rm in}^3 
+ 129 h_{\rm in}^4)}{3969} \right]^{-1},
\label{padepsia}
 \\
& &_{} 
\mPsi_{1 \, \mbox{\scriptsize Pad\'e}(2,1)}=
\mp \frac{a (1 + h_{\rm in}) q_1}{21} \times
\nonumber\\
&& \quad
\left[567 (1 - h_{\rm in}) \right.
\left.\mp a (80 - 80 h_{\rm in} - 31 h_{\rm in}^2 
+ 129 h_{\rm in}^3)\right] \times
\nonumber\\
&& \quad
\left[81 (1 - h_{\rm in})\right.
\left. \mp a (23 - 23 h_{\rm in} - 
16 h_{\rm in}^2 + 30 h_{\rm in}^3) \right]^{-1},
\label{padepsib}
\end{eqnarray}
where we employed Pad\'e approximation of type-(1,2) and type-(2,1) in
Eqs.~(\ref{padepsia}) and (\ref{padepsib}), respectively.

We show the result of these Pad\'e-prescribed PZA and PPZA (Pad\'ePZA,
Pad\'ePPZA hereafter) in Figs.~2 and 3 as well as other non-linear
approximations. Since the relative accuracy of various approximations
is almost the same for all values of $\delta_{\rm true}$, we have
decided to fix $\delta_{\rm true}$ and show the accuracy of
approximations for various values of $\alpha_{\rm in3}/\alpha_{\rm
in1}$ at once.  In Fig.~2, the density contrasts of various
approximations in spheroidal collapse of positive perturbations are
shown.  The horizontal axis represents the initial axis-ratio
$\alpha_{\rm in3}/\alpha_{\rm in1}$; oblate configuration for
$\alpha_{\rm in3}/\alpha_{\rm in1}<1$, and prolate configuration for
$\alpha_{\rm in3}/\alpha_{\rm in1}>1$.  The case $\alpha_{\rm
in3}/\alpha_{\rm in1}=1$ corresponds to the spherical configuration
shown in Fig.~1.  The vertical axis in Fig.~2 represents the ratio of
the density contrasts $\delta_{\rm approx}/\delta_{\rm true}$
evaluated at $\delta_{\rm true}=4$ well within the non-linear
regime. Both the axis ratio and density ratio are in logarithmic scale
in this figure.  In Fig.~3, we plotted the same graph as Fig.~2 but
for negative perturbations.  The density contrast $\delta_{\rm
approx}/\delta_{\rm true}$ in the negative perturbation model is
evaluated at $\delta_{\rm true}=-0.6$.

We observe from these figures the following facts:
\begin{itemize}
\item[(i)] Pad\'ePPZA constantly yields the best precision among
  various approximation scheme for all initial axes-ratio $\alpha_{\rm
  in~3}/\alpha_{\rm in~1}$ in both positive and negative spheroidal
  perturbations.
\item[(ii)] The Pad\'e-prescription results in much dramatic
  improvement in precision in the negative perturbations than positive
  perturbations
\item[(iii)] All the graph seems to converge to the exact solution in
  the oblate limit.
\end{itemize}
The above facts (i) and (ii) demonstrate the excellence of the Pad\'e\ 
prescription for ZTA.  Especially the negative perturbation case is
remarkable.  For example, the Pad\'e prescription improves the PPZA
about factor six for $\delta_{\rm true}=-0.6$.  The last fact (iii)
simply reflects that Pad\'e approximations as well as Zel'dovich-type
approximations have the one-dimensional-exact property.  Actually the
limit of one-dimensional collapse is achieved by setting $h_{\rm in}
\rightarrow -1$.  In this limit, the Eq.~(\ref{eq:mm2}) reduces
to the exact solution, $\delta = \pm a/(1 \mp a)$.  Therefore the
Pad\'e approximation in the present model also has the
one-dimensional-exact property as ZTA.

\begin{minipage}{8.5cm}
\begin{figure}
\begin{center}
   \leavevmode\psfig{figure=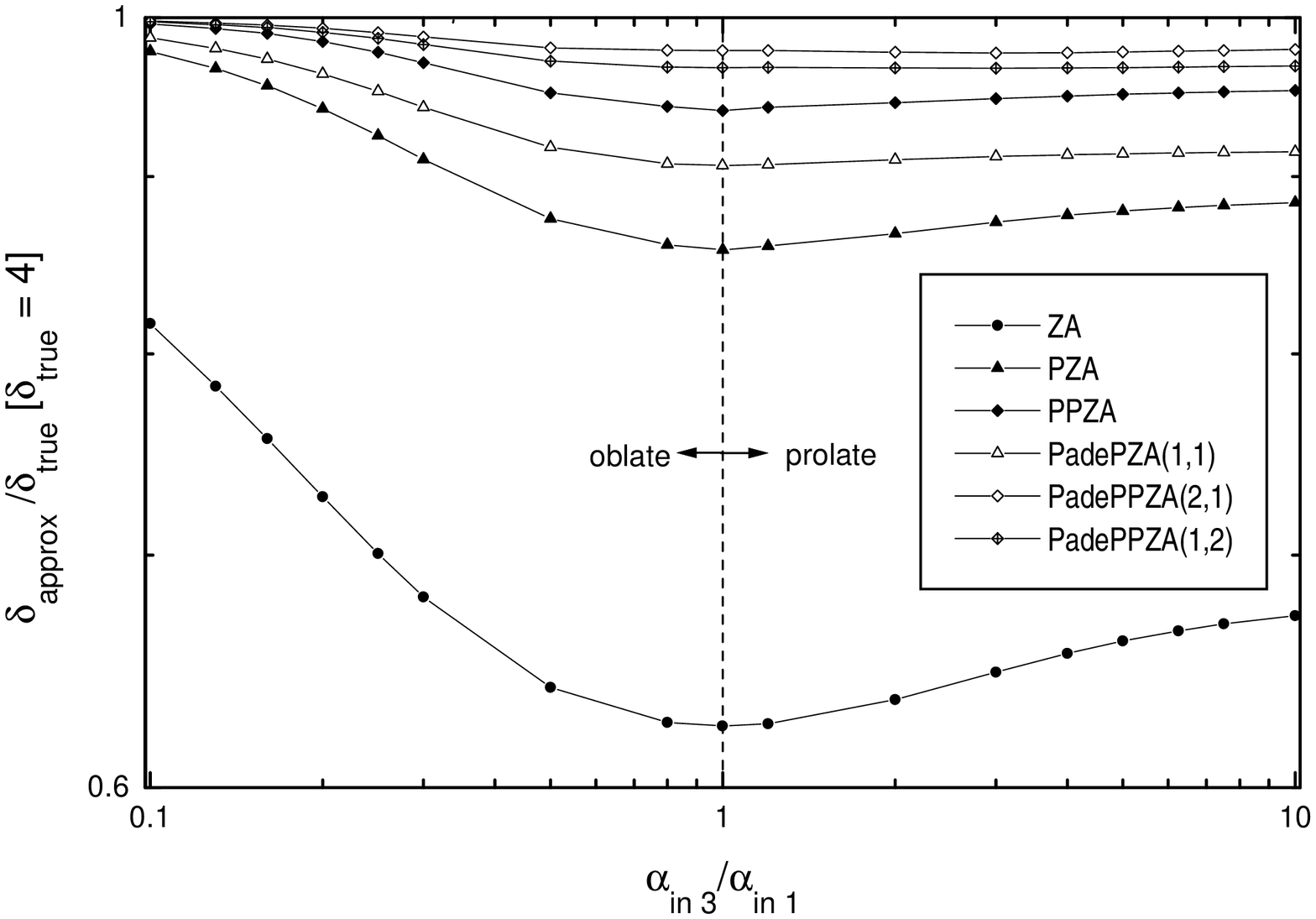,width=8.5cm,angle=270}
\end{center}
\caption{Density contrasts of various approximations in
spheroidal collapse of positive perturbations.  The horizontal axis
represents the initial axis-ratio $\alpha_{\rm in3}/\alpha_{\rm in1}$
and the vertical axis represents the ratio of density contrast
$\delta_{\rm approx}/\delta_{\rm true}$ evaluated at $\delta_{\rm
true}=4$.  Both the axis ratio and density ratio are in logarithmic
scale in this figure.}
\end{figure}
\end{minipage}

\begin{minipage}{8.5cm}
\begin{figure}
\begin{center}
   \leavevmode\psfig{figure=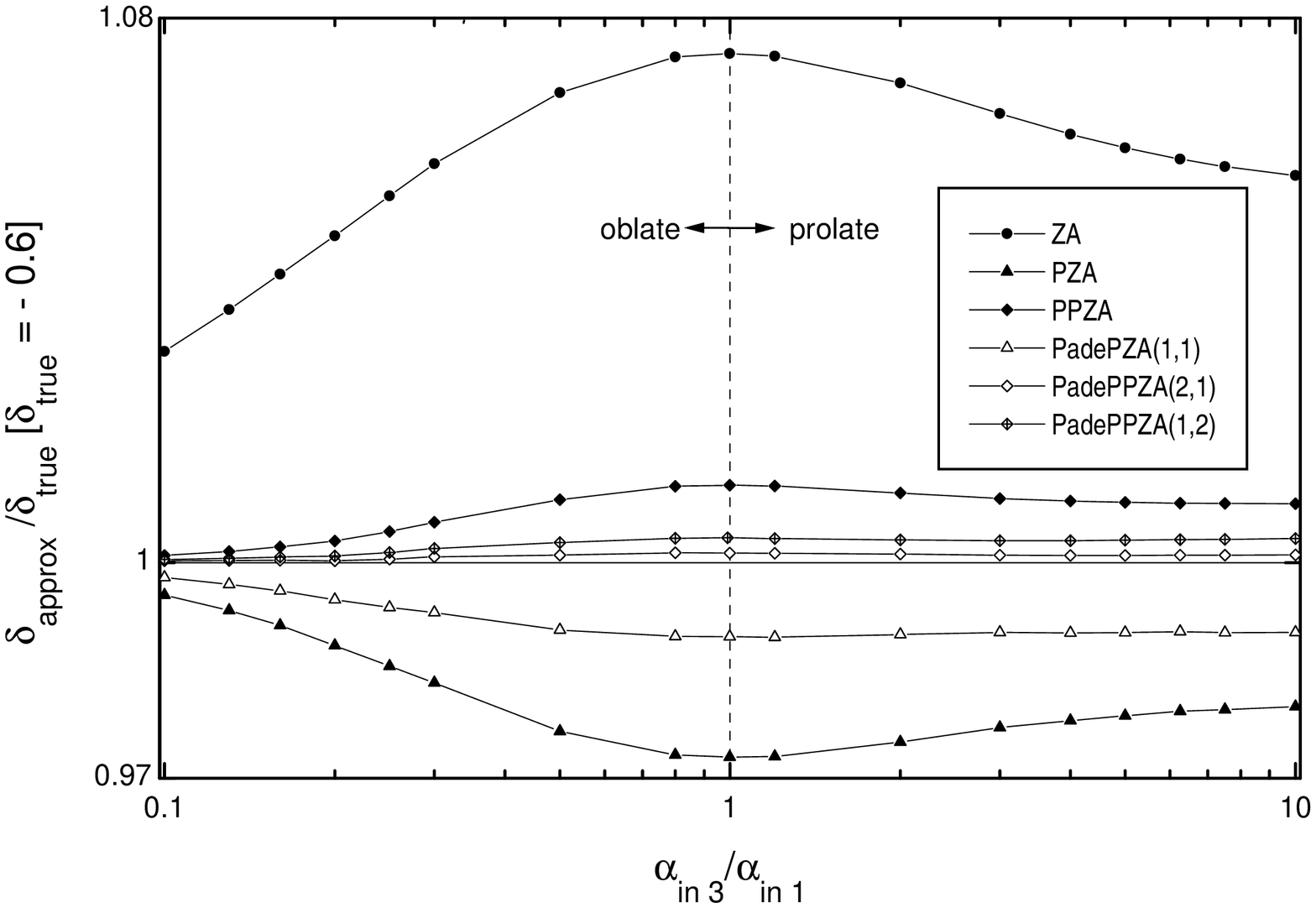,width=8.5cm,angle=270}
\end{center}
\caption{The same graph as Fig. 2 but for negative
perturbations.  The vertical axis represents the ratio of density
contrast $\delta_{\rm approx}/\delta_{\rm true}$ evaluated at
$\delta_{\rm true}=-0.6$.}
\end{figure}
\end{minipage}
\vspace{5mm}

Now we conclude our work.  First we focused on the issue that ZTA and
Pad\'e approximations are actually the same scheme or not.  By using
spheroidal perturbation models, we found that these two types of
approximations are different independent scheme with each other
despite the similar appearance.  Then, based on this fact, we proposed
the Pad\'e prescription for the ZTA. We found this Pad\'e prescription
appreciably increases the accuracy of ZTA.

This Pad\'e-prescribed ZTA may shed light on the practical
calculations on the evolution of density perturbations in the
Universe.

The Pad\'e approximation for PPZA we used is not the pure original
form of the scheme.  Actually we Pad\'e-prescribed the part of
denominator of the density perturbations $\delta$ in
Eq.~(\ref{yuragi}).  This reminds us the continued fraction
approximation.  We hope these Pad\'e and continued fraction
approximation scheme may reveal the true validity of the ZTA in the
future.

\end{multicols}


\begin{references}
\bibitem{foot1} The coefficients $A_i$ automatically satisfy the
  constraint $\sum A_i = 2$, which follows from Poisson
  Eq.~(\ref{eq3}).
\end{references}
\end{document}